\documentclass[conference,12pt,onecolumn,british,oneside,a4paper]{IEEEtran}
\IEEEoverridecommandlockouts
\usepackage{amsmath}
\usepackage{amssymb}
\usepackage{graphicx}
\usepackage{todonotes}
\usepackage{textcomp}

\usepackage{mathptmx}
\usepackage{tabularx}
\usepackage{ragged2e}
\usepackage{amsmath}
\usepackage[british]{babel}

\DeclareGraphicsExtensions{.eps,.pdf,.jpeg,.png}
\usepackage[caption=false,font=footnotesize]{subfig}
\usepackage{color}

\usepackage{fancyhdr}

\DeclareMathOperator{\erf}{erf}
\newcommand{\mult}{\,}

\begin{document}
	
	\title{An AI-driven Malfunction Detection Concept for NFV Instances in 5G}
	\author{\IEEEauthorblockN{Julian Ahrens\IEEEauthorrefmark{1},
			Mathias Strufe\IEEEauthorrefmark{1}, Lia Ahrens\IEEEauthorrefmark{1} and Hans Dieter Schotten\IEEEauthorrefmark{2}\IEEEauthorrefmark{1}}
		\IEEEauthorblockA{\IEEEauthorrefmark{1}German Research Centre for Artificial Intelligence (DFKI)\\Trippstadter Street 122,  Kaiserslautern, 67663 Germany\\
			Emails: \{julian.ahrens, mathias.strufe, lia.ahrens, hans.schotten\}@dfki.de  }
		\IEEEauthorblockA{\IEEEauthorrefmark{2}Institute for Wireless Communication and Navigation,
			University of Kaiserslautern\\Building 11, Paul-Ehrlich Street, Kaiserslautern, 67663 Germany\\
		}
		\thanks{$^\ast$This work was supported by the European Union's Horizon 2020 Programme under the 5G-PPP project: \emph{Framework for Self-Organized Network Management in Virtualized and Software Defined Networks}\,(SELFNET) with Grant no. \emph{H2020-ICT-2014-2/671672}.}
	}
	\maketitle

\begin{abstract}
	Efficient network management is one of the key challenges of the constantly growing and increasingly complex wide area networks (WAN). The paradigm shift towards virtualized (NFV) and software defined networks (SDN) in the next generation of mobile networks (5G), as well as the latest scientific insights in the field of Artificial Intelligence (AI) enable the transition from manually managed networks nowadays to fully autonomic and dynamic self-organized networks (SON). This helps to meet the KPIs and reduce at the same time operational costs (OPEX).
	
	In this paper, an AI driven concept is presented for the malfunction detection in NFV applications with the help of semi-supervised learning. For this purpose, a profile of the application under test is created. This profile then is used as a reference to detect abnormal behaviour. For example, if there is a bug in the updated version of the app, it is now possible to react autonomously and roll-back the NFV app to a previous version in order to avoid network outages.
\end{abstract}

\section{Motivation}

As of today, maintenance and servicing of permanent growing mobile networks require manual intervention of qualified network engineers to ensure a constant high level of service quality which is very time and cost consuming. Operators need to locate and mitigate different type of problems in the network, such as hardware faults, link failures, performance optimization and security attacks, to only name a few.
The European Commission (EC) and others highlighted already in 2014 that mobile operators spending three times operational expenditures (OPEX) than capital expenditures (CAPEX) \cite{Ref:EU}\cite{Ref:Aviat}.

With the emerging Fifth Generation (5G) \cite{Ref:NGMN} and therefore gaining heterogeneous and complex networks this numbers will increase further. The network function virtualization (NFV) \cite{Ref:NFV} and software defined network (SDN) \cite{Ref:SDN} principles of the future core networks allow more flexibility but also enable the option to automate many of that maintenance and management tasks.
The {EU} {H2020} {SELFNET} \cite{Ref:SELFNET}\cite{Ref:SELFNET2} project is addressing these challenges and developing a self-organized 5G network management framework through virtualized and software defined networks to support these new technologies and reduce OPEX.

The SELFNET framework mainly consists of: %\begin{itemize}[noitemsep,topsep=0pt]
1.) SDN/NFV sensors - that extract metrics from the network infrastructure,
2.) Aggregation Layer - for preprocessing and storage of the collected data,
3.) Autonomic Engine, which is divided into 
%	\begin{itemize}[noitemsep,topsep=0pt]
a.) Rule based and
b.) AI based decision engine, 
%	\end{itemize}
4.) Action Enforcer and Orchestrator - that implement the Decision from the Autonomic Management,
5.) SDN/NFV actuators - e.g. FlowControl Agent;
%\end{itemize}

The SELFNET consortium focuses on three main use-cases: SELF-PROTECTION, with its capabilities against distributed cyber-attacks
SELF-HEALING that handles system failures and SELF-OPTIMIZATION to improve dynamically the performance of the network and the QoE of the users.

While self-optimization and self-protection already carried out and described in \cite{Ref:ICCC2017}\cite{Ref:PIMRC2017} and \cite{Ref:Botnet}, we focus in this paper on the self-healing capabilities.

The paper is organized as follows: Section 2 gives an overview of the intelligence concept and used methods. The set-up of the test-bed for experimentation and evaluation of the Intelligence is presented in Section 3. Finally Section 4 concludes this paper.

\section{Dataset Generation and Evaluation}
In this section the set-up of the test-bed and methods to create the training dataset and evaluation of the proposed Intelligence concept is described. 

\subsection{Test Set-up}

In this first iteration of NFV self-healing experimentation, the following virtualized test environment was created to generate the initial learning dataset.

\begin{figure*}[!ht]
	\centering
	\includegraphics[width=0.95\textwidth]{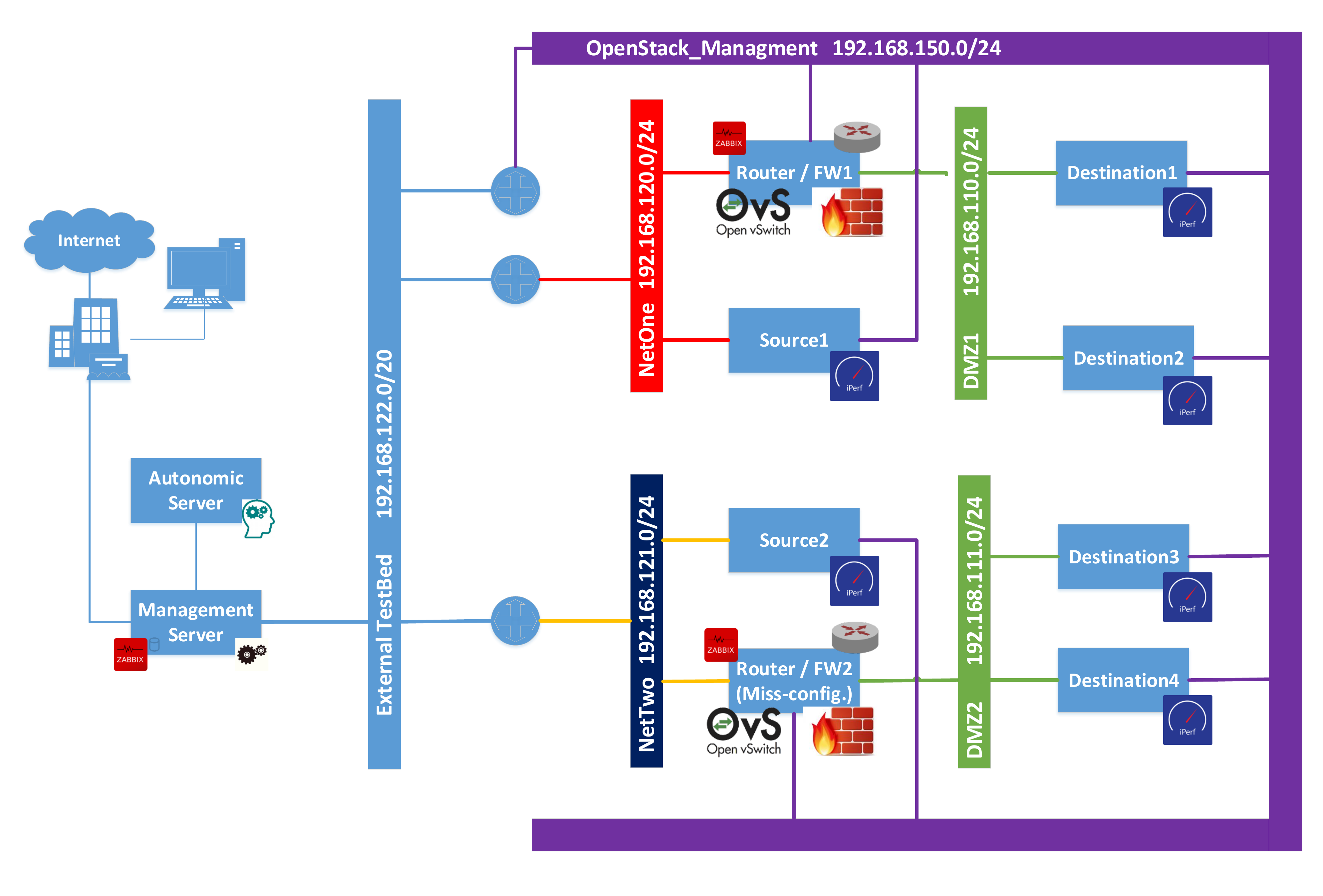}
	\caption{Self-Healing Test environment.}\label{Fig:SHTopo}
\end{figure*}

Basically it consists of six networks and eight virtual machines. 
The external network is the connection between the network under test (on the right side) and the LAB network where also the Management and Autonomic Server is located, see Figure \ref{Fig:SHTopo}.
The Management Server serves as Gateway to the Internet and simultaneously as centralized storage for the monitored data with the help of the open-source tool Zabbix \cite{Ref:Zabbix}. 
Since the foundation for the virtualization is OpenStack \cite{Ref:OS}, the OpenStack management network is used to maintaining the virtual machines as well as have metric collection strictly separated from the network under test.
The virtual machine called Source, connected to the NetOne network, acts as Web Server, in this first version running the Iperf3 \cite{Ref:iperf} tool in server mode that generates traffic to the two virtual machines in the DMZ1 network, the so called Destination VMs, that acts as user and running as well the Iperf3 toll in client mode. 
The Traffic needs to pass the Firewall between, that connect NetOne and DMZ1. This Firewall NFV is running in general the OpenVSwitch \cite{Ref:OVS} with a WebInterface as FrontEnd to inject new rules and was original developed in the 5GPPP Charisma Project \cite{Ref:Charisma}.
These constellation is duplicated in the NetworkTwo and DMZ2, with the difference that this Firewall2 is miss configured, to have a direct reference.

On both Firewalls a Zabbix agent is running to collect metrics, such as CPU, memory, harddisk usage and traffic in/out. These basic metrics will be used to create the profile of the NFV to find out if it is working in a proper manner. 

\begin{figure*}[!ht]
	\centering
	\includegraphics[width=0.8\textwidth]{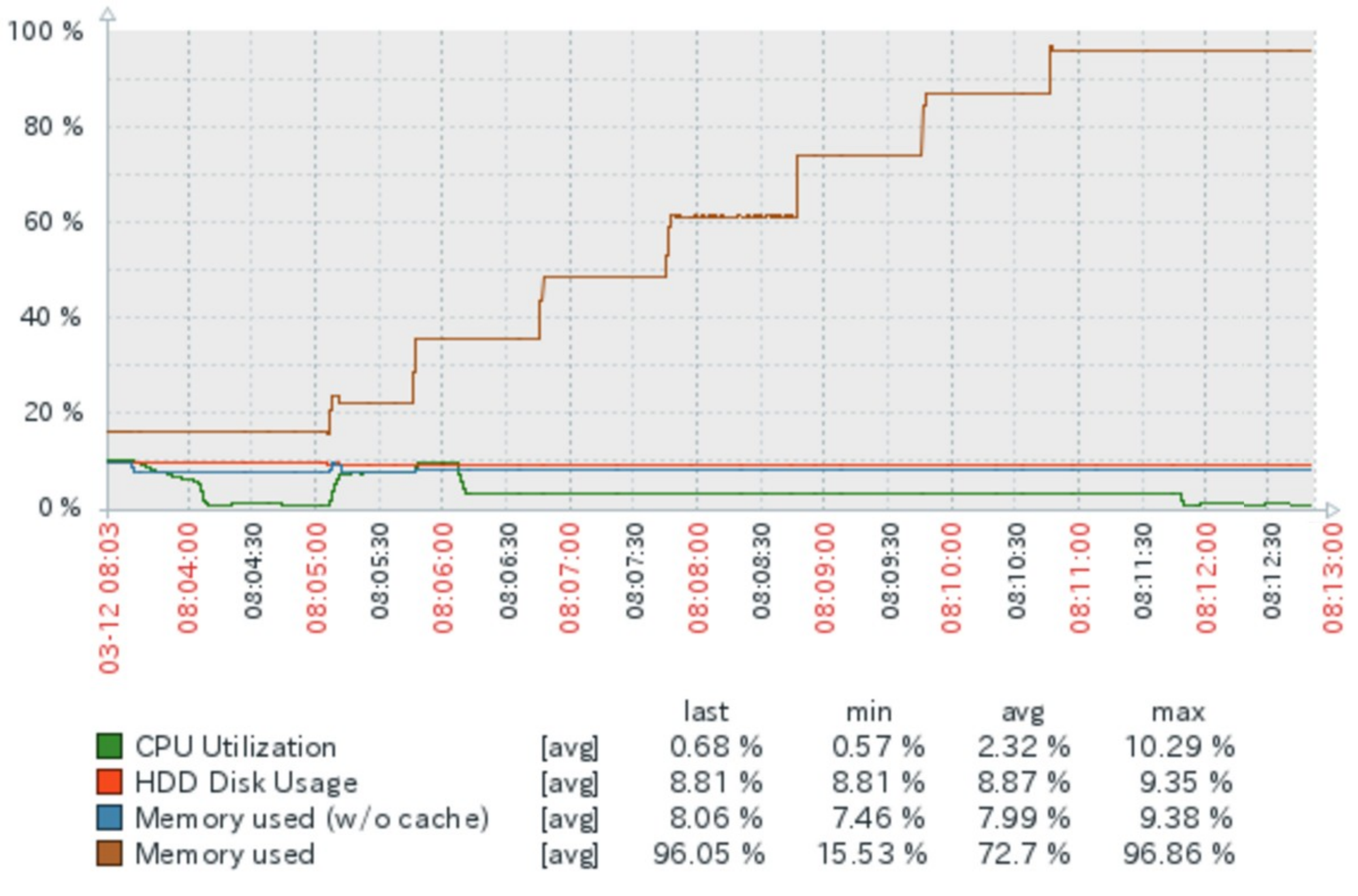}
	\caption{Firewall Metrics - CPU, Harddisk and Memory (with and without cache) usage.}\label{Fig:MEMGraph}
\end{figure*}

The miss configuration in this example is a fictive memory leak, that is in this test generated with the stress-ng tool, a to load and stress a computer systems. It is started with the parameters to allocate every minute 200~MB of memory and keep it. Since the NFV only has 1~GB memory in total, it only needs a few minutes to occupy almost 100~\%, see Figure~\ref{Fig:MEMGraph}.

\section{Artificial Intelligence}

\subsection{Approach}
We employ a semi-supervised learning approach.
The training process uses a dataset which contains measurements taken during known good operation of the VNF.
This training set can be extended during operation:
Whenever an anomaly is detected, the problem is reported.
The operator can then inspect the operation of the VNF and, in case of the problem report being a false positive, can opt to add the corresponding data to the training set thereby enabling the detector to know about this newly occurring region of regular operation.

\subsection{Techniques}

\subsubsection{Pre-processing}
We employ a specialized normalization method, which transforms each feature into a random variable following a particular normal distribution.
This is achieved by computing for each feature $X_{k}$ the empirical distribution function $\hat{F}_{k}$ accross the training data and using a (continuous, strictly increasing) modification $F_{k}$ of this function as an estimator for the distribution of $X_{k}$.
The data is then transformed on a per-feature basis by applying the transformation
\begin{equation}
	X_{k} \mapsto \Phi^{-1}(F_{k}(X_{k}))
\end{equation}
where $\Phi^{-1}$ denotes the quantile function of the standard normal distribution which can be expressed in terms of the inverse error function $\erf^{-1}$ as
\begin{equation}
	\Phi^{-1}(p)
	= \sqrt{2} \mult \erf^{-1}(2 \mult p - 1)
\end{equation}

The above mentioned modification of the empirical distribution function is performed in the following way:
First the steps of the empirical distribution function are replaced by linear segments continuously joining together at the ends and sloped anti-proportionally to the step length.
Then the cumulative distribution function of Gaussian random variable is mixed into the resulting function in order to ensure strictly monotonic behaviour even in regions outside of the data given by the training set.

\subsubsection{Autoencoder}
The actual outlier detection is performed using an ensemble of autoencoders $E_{l}$ of different shapes learning at different rates.
An autoencoder is a function $E_{l}$ mapping values $x$ to approximations $\tilde{x}$ of $x$.
Their parameters $p$ are adjusted by a stochastic gradient descent type learning algorithm to achieve the best possible approximation.
The goodness of fit is measured using the square of the standard Euclidean distance $\lVert \tilde{x} - x \rVert_{2}^2$.
The training algorithm then optimizes the parameters $p_{l}$ to minimize the cost
\begin{equation}
	\sum_{x \in \mathcal{X}} c(x, p_{l})
	= \sum_{x \in \mathcal{X}} \lVert E_{l}(x, p_{l}) - x \rVert_{2}^2
\end{equation}
where $\mathcal{X}$ denotes the training dataset.
This optimization is performed for each individual autoencoder $E_{l}$.

\subsubsection{Thresholding}
The cost functions of each of the autoencoders are used as estimators of the degree of anomality of the data.
After training, the autoencoders are evaluated on a set of validation data.

The ensemble is then reduced to only the best $m$ encoders.
For each incoming datapoint, the costs w.r.t.\ these remaining encoders are computed.

If the cost w.r.t.\ an encoder $E_{l}$ is more than $\beta$ times the average cost of this particular encoder on the validation dataset, the datapoint is marked as suspicuous.

If more than $\alpha$ of the encoders mark the datapoint as suspicuous, it is considered an anomaly.

\section{Conclusion}
In this paper, we presented a semi-supervised learning concept for monitoring NFV Instances and perform malfunction detection. The set-up of the test environment for the dataset creation was described as well as the methods for the machine learning approach. In a next step we will improve the algorithms in terms of robustness and extend the test environment to simulate more malfunctions and enlarge the environment to a more complex version.

\bibliography{Ref_ITG} 
\bibliographystyle{IEEEtran}

\end{document}